\documentclass[preprint,aps,amsmath]{revtex4}

\DeclareMathOperator{\tr}{tr}
\newcommand{\frct}[2]{{\textstyle\frac{#1}{#2}}}
\renewcommand{\vec}[1]{{\rm\bf #1}}
\newcommand{\de}{\,{\rm d}}
\newcommand{\te}{{\rm T}}
\newcommand{\si}{\sigma}

\newcommand{\ep}{\epsilon}
\newcommand{\epv}{\varepsilon}
\newcommand{\al}{\alpha}
\newcommand{\ka}{\kappa}
\newcommand{\la}{\lambda}

\begin{document}

\title{A new parametrizable model of molecular electronic structure}

\author{Dimitri N. Laikov}
\email[E-Mail: ]{laikov@rad.chem.msu.ru}
\homepage[Homepage: ]{http://rad.chem.msu.ru/~laikov/}
\affiliation{Chemistry Department, Moscow State University,
119992 Moscow, Russia}

\date{\today}

\begin{abstract}
A new electronic structure model is developed
in which the ground state energy of a molecular system
is given by a Hartree-Fock-like expression
with parametrized one- and two-electron integrals
over an extended (minimal + polarization) set
of orthogonalized atom-centered basis functions,
the variational equations being solved formally within the minimal basis
but the effect of polarization functions being included in the spirit
of second-order perturbation theory.
It is designed to yield good dipole polarizabilities
and improved intermolecular potentials with dispersion terms.
The molecular integrals include up to three-center one-electron
and two-center two-electron terms,
all in simple analytical forms.
A method to extract the effective one-electron Hamiltonian
of nonlocal-exchange Kohn-Sham theory from the coupled-cluster one-electron
density matrix is designed and used
to get its matrix representation in a molecule-intrinsic minimal basis
as an input to the paramtrization procedure --
making a direct link to the correlated wavefunction theory.
The model has been trained for 15 elements (H, Li--F, Na--Cl, 720 parameters)
on a set of 5581 molecules
(including ions, transition states, and weakly-bound complexes)
whose first- and second-order properties were computed
by the coupled-cluster theory as a reference,
and a good agreement is seen.
The model looks promising for the study of large molecular systems,
it is believed to be an important step forward
from the traditional semiempirical models
towards higher accuracy at nearly as low a computational cost.
\end{abstract}

\maketitle

\section{Introduction}

Atomistic computer simulations of complex chemical systems and materials
at quantitative level are a great challenge for modern science:
not only a high accuracy of computed potential energy surfaces
is needed for systems with many atoms,
but also a higher speed of computation for a thorough sampling
of the configurational space.
Molecular mechanical force fields pioneered~\cite{LW68} 44 years ago
are still almost the only practical method of calculation in many fields
thanks to their very high speed and despite their well-known limitations.
With their fixed bonding topology, they are designed first of all
for conformational studies where nonbonding interactions play the main role,
their treatment of electrostatics may be as sophisticated
as to account for polarization effects~\cite{WL76}.
Extensions to treat one (or few) chemical reaction steps~\cite{WW80}
need careful parameter adjustment for each active center studied.
General reactive force fields~\cite{B90,DDLG01}
with geometry-dependent bond orders and atomic charges
seem to be a logical next step and already remind
of models within the Hohenberg-Kohn~\cite{HK64} density functional theory (DFT).
Our own experience with this kind of models lead us to believe
that instead of designing more and more complicated
charge and bond order functionals it should be much easier
to incorporate at least a single matrix diagonalization
of an effective one-electron Hamiltonian~\cite{H63} into the model
(likely in a linear-scaling fashion~\cite{LNV93,PM98}),
just like the mainstream DFT took the Kohn-Sham path~\cite{KS65},
and thus we escape the force field and
enter the realm of molecular quantum mechanics.

At the other extreme, rigorous wavefunction methods give useful results
starting with the second-order many-body perturbation theory (MP2)~\cite{MP34},
but much better with the coupled-cluster theory~\cite{PCS72}
with single and double substitutions (CCSD)~\cite{PB82}
or -- as the golden standard -- with the further perturbative account
of triple substitutions (CCSD(T))~\cite{RTPH89}.
Their fifth, sixth, and seventh power scaling of computational cost
with the system size and huge prefactors, as well as the slow convergence
of the computed properties with the basis set size,
make them hopelessly slow for large molecules,
but they are indispensable for getting small-molecule reference data
for training and testing all kinds of models.

Kohn-Sham DFT on its way from local~\cite{KS65} to generalized-gradient
approximations~\cite{B88,PBE96},
with further inclusion of a fraction~\cite{B93}
or, much better, the full long-range part~\cite{ITYH01} of the nonlocal exchange
and the dispersion tail~\cite{LASCL95,ALL96},
has grown into a rather accurate electronic structure model
with favorable system-size scaling properties.
In a standard implementation, localized atom-centered basis functions
are used to solve the self-consistent field (SCF) equations
(plane wave techniques~\cite{CP85} are limited to the less accurate
pure density functionals and cannot work as fast with the nonlocal exchange),
the analytical evaluation of two-electron Coulomb integrals
and the numerical integration of exchange-correlation terms
are the bottleneck for smaller system sizes but can be made linear scaling
starting from around 1000 atoms,
in that limit we estimate the computation of DFT energy functional to be
about $10^6$ to $10^8$ times slower than the most sophisticated
polarizable force field energy and gradient evaluation.
Density-fitting techniques~\cite{DCS79,L97} can speed up the calculation
of these terms by up to 100 times, but only for the pure DFT.
Pseudospectral methods~\cite{F88,MCBRF00} can show up to 100 times speed-up
also for the nonlocal-exchange DFT. Even if the integrals were for free,
there would be another serious bottleneck in the linear algebra
of SCF equations.
A smallest meaningful atomic basis set with 5 functions for H and 14 or 18
for Li--Ne would yield little to no sparsity of the density matrix
for typical three-dimensional molecules with up to 3000 atoms~\cite{RRS08}
and around 30000 basis functions, so about $10^{13}$
floating-point multiplications and additions would be needed
to do one matrix-multiply! In this regime, one SCF energy calculation
can hardly take much less than one day
even on a modern high-performance parallel computer.

The only way for the electronic structure theory to compete
with the force field methods lies in the use
of a finite-dimensional model Hamiltonian defined by its matrix elements
in a minimal atomic basis representation.
This is an alternative strategy with the Kohn-Sham DFT --
instead of defining the universal density functional in some limited
(approximate, parametrized) form and then computing the arising integrals
rigorously at each molecular geometry, these molecular integrals themselves
can be directly modeled as the functions of atomic coordinates.
One may note that the Coulomb potential of the nuclei in molecules
is a very special class of external potentials for a system of electrons,
and the widely-used density functionals are not as universal as thought,
being often fitted~\cite{B97} to molecular data.
The minimal basis representation is not a limitation --
not a fixed free-atom but a molecule-adapted set
of (deformed) atomic functions is implied.
In our earlier work~\cite{L11} we have shown how to extract
such a basis set from an accurate (or exact) solution
of Kohn-Sham equations at a given molecular geometry
in a unique and molecule-intrinsic way.
The effective local potential of Kohn-Sham theory
can in turn be extracted from the correlated wavefunction theory
by a stable numerical procedure~\cite{ZMP94},
in Section~\ref{sec:FD} of this work we derive its analog
for the case of nonlocal exchange.
(After we had done ours, we were made aware~\cite{A11} of a parallel work
on the effective valence shell Hamiltonians~\cite{F72,IF76,F83,MF96}
but for a fully \textit{correlated} -- and not SCF -- treatment
of the ground and excited states).
Now that the effective minimal-basis matrix elements of the model Hamiltonian
have the first-principles foundation and can be computed numerically,
they can be used to guide the work on their approximations.

Into this category fall the semiempirical electronic structure models
based on the neglect of diatomic differential overlap (NDDO) formalism
that have had nearly half a century of conservative evolution.
Conceived as the simplest approximations~\cite{PSS63}
for valence-only minimal-basis SCF calculations
with most multicenter integrals set to zero,
they became a breakthrough in computational chemistry~\cite{D75}
when a systematic parametrization~\cite{DT77,NJ80,DZHS85,S89,FGZ87,S07}
to fit experimental molecular data
turned them into predictive phenomenological models.
As the growing computer power allowed the DFT, MP2, and other rigorous
methods to be applied to chemically-interesting systems
and the standards of accuracy tightened,
the semiempirical models began to lose the game.
Some limited orthogonalization corrections were studied~\cite{KT93,WT00}
but lead only to a limited improvement.
Poor intermolecular potentials, especially for hydrogen-bonding,
were tried to be cured~\cite{NH07,TT08}
by simple diatomic dispersion corrections~\cite{APS77}
and further by triatomic corrections~\cite{RFSH98,K10}
in the spirit of molecular mechanics --
by adding the terms that depend on the atomic coordinates only
and not on the electronic state.
A proper treatment of polarizability within the minimal-basis formalism
is not straightforward,
one solution~\cite{RMW82} is to add polarization functions into the basis
but this slows down the calculations.
A more attractive way was found~\cite{GY05,CSG08}
in which a polarization term borrowed from molecular mechanics
is added on top of the SCF equations.
With all these recent developments, the desired improvement
in accuracy of the semiempirical NDDO models for many applications
is still not achieved.

Here we report our new parametrizable electronic structure model
that evolves along a different line.
We consider an extended atomic valence basis set
that has radial and angular polarization functions
added to the minimal set on each atom,
and then we design a new two-layer SCF method (Section~\ref{sec:E})
in which the variational equations are defined within the minimal subspace
and the contribution from the polarization subspace is included
in the spirit of second-order perturbation theory.
It should be noted that our SCF method is fairly different in many ways
from other known dual-basis techniques~\cite{LH04,SH07,DGG09}.
The parametrizable molecular integrals of our model (Section~\ref{sec:MI})
have more general functional forms and include new terms not seen
in the traditional semiempirical NDDO models.
We add the long-range dispersion corrections into the two-electron
integrals -- a more meaningful treatment of the two-electron correlation effect
that depends on the electronic state.
Our model also treats the molecular polarizability in a natural
and self-interaction-free way.
The parametrization procedure (Section~\ref{sec:P}) we use to train our model
has a number of new terms added into the optimization process:
the effective one-electron Hamiltonian matrices extracted (Section~\ref{sec:FD}
and~\cite{L11}) from the correlated wavefunction theory,
the molecular dispersion coefficients (Section~\ref{sec:C6})
and the molecular electrostatic potentials on the surfaces (Section~\ref{sec:SV}),
as well as the second-order response properties (force constant matrices,
atomic polar tensors, and dipole polarizabilities) --
all these values coming from the high-level (coupled-cluster) calculations
are required to be reproduced (in the least-squares sense)
by the parametrized SCF model.

All this methodology defines our model up to the values of parameters
that have to be optimized on a set of reference molecular data.
Here we report (Section~\ref{sec:calc}) a preliminary parametrization
for 15 elements H, Li--F, Na--Cl on a very diverse set of molecular structures
to assess the worst-case performance of the model --
it stands the test, seems to be already usable and useful for some applications,
and awaits a future extension to heavier elements.

\section{Theory}
\label{sec:theory}

\subsection{Energy expression}
\label{sec:E}

In our model, the total energy of a molecular system
with $K$ nuclei at positions $\{\vec{r}_k\}$, $k=1,\dots,K$,
in an external field $v (r)$
can be written as a sum of three terms
\begin{equation}
\label{eq:E}
E = E_0 + E_1 + E_2,
\end{equation}
the one independent of the valence electronic structure
\begin{equation}
E_0 =
  \sum\limits_k e_k
+ \sum\limits_{k<k'} q_k q_{k'} |\vec{r}_k - \vec{r}_{k'}|^{-1} 
+ \sum\limits_k q_k v (\vec{r}_k)
\end{equation}
being the sum of core electron energies $e_k$,
the repulsion of atomic cores with charges $\{q_k\}$
and their interaction with the external field;
the Hartree-Fock energy within the minimal atomic basis
\begin{equation}
E_1 = \frct12 \sum\limits_{\mu\nu\si}
\left( H_{\mu\nu} + F_{\mu\nu\si} \right) D_{\mu\nu\si},
\end{equation}
\begin{equation}
\label{eq:F0}
F_{\mu\nu\si} = H_{\mu\nu}
+ \sum\limits_{\mu'\nu'\si'} R_{\mu\nu\si\mu'\nu'\si'} D_{\mu'\nu'\si'},
\end{equation}
the indices $\mu,\nu=1,\dots,M$ running over the minimal set,
the spin label $\si=\pm\frct12$;
and the second-oder term
\begin{equation}
E_2 = \sum\limits_{\mu\nu\si}
F^{(2)}_{\mu\nu\si} D_{\mu\nu\si},
\end{equation}
\begin{equation}
\label{eq:F2}
F^{(2)}_{\mu\nu\si} =
\sum\limits_\al \left(
 F_{\al\mu\si} X_{\al\nu\si} \bar{s}_{\nu\si} +
 F_{\al\nu\si} X_{\al\mu\si} \bar{s}_{\mu\si}
\right),
\end{equation}
\begin{equation}
\label{eq:xam}
X_{\al\mu\si} = \frac{F_{\al\mu\si}}{\epv_\mu - \epv_\al},
\end{equation}
\begin{equation}
\label{eq:F1}
F_{\al\nu\si} = H_{\al\nu}
+ \sum\limits_{\mu'\nu'\si'} R_{\al\nu\si\mu'\nu'\si'} D_{\mu'\nu'\si'},
\end{equation}
that accounts for the effect of polarization functions
(labeled by $\al=M+1,\dots,N$) in a perturbative way.
The one-electron density matrices
\begin{equation}
D_{\mu\nu\si} = \sum\limits_{i=1}^{N_\si} C_{\mu i\si} C_{\nu i\si}
\end{equation}
with $N_\si$ electrons for each spin are formed
from the orthogonal coefficient matrices
\begin{equation}
\label{eq:Corth}
\sum\limits_\mu C_{\mu i\si} C_{\mu j\si} = \delta_{ij}.
\end{equation}
The orthogonalized atomic valence basis functions
\begin{equation}
\label{eq:phi}
\phi_\mu = \phi_{m_\mu l_\mu n_\mu k_\mu}
\end{equation}
can also be labeled by their principal $n$, angular $l$, and azimuthal $m$
quantum numbers and atomic centers $k$, for each $l$ no more than one $n$ value
is used in either minimal or polarization set.
The one-electron integral matrix is a sum
\begin{equation}
\label{eq:H1}
H_{\mu\nu} = T_{\mu\nu} + \sum\limits_{k'} V_{\mu\nu k'} + V_{\mu\nu}^{(v)}
\end{equation}
of kinetic energy, effective core potential, and external field terms.
The spin-dependent two-electron integrals
\begin{equation}
R_{\mu\nu\si\mu'\nu'\si'} =
R_{\mu\nu\mu'\nu'} - \delta_{\si\si'} R_{\mu'\nu\mu\nu'}
\end{equation}
are built from the spinless Coulomb repulsion integrals
by the proper antisymmetrization.
The values $\epv_\mu < 0$ and $\epv_\al > 0$ in the denominator
of Eq.~(\ref{eq:xam}) are taken as atomic constants
and play the role of diagonal energies of the zeroth-order problem
of perturbation theory, their signs imply that $E_2\leq0$.
Had we had $\bar{s}_{\mu\si}\equiv 1$ in Eq.~(\ref{eq:F2})
as we did in an earlier model, the total energy $E$ would be a cubic function
of the density matrix, whereas the Hartree-Fock energy $E_1$ alone is quadratic --
the perturbative inclusion of higher-order basis set effects
is leading to the appearance of effective three-electron integrals.
Moreover, $E_2$ would be quadratic in the external field -- as needed
for the proper account of polarizability within a formally minimal-basis
treatment.
Our experience has shown, however, that the simplest choice
of $\bar{s}_{\mu\si}\equiv 1$
often leads to a collapse of $E_2$ when atoms in a molecule get crowded --
we have overcome this problem using the damping factors
\begin{equation}
s_{\mu\si} = \left(1 + \sum\limits_\al X^2_{\al\mu\si} \right)^{-1}
\end{equation}
that need to be spherically averaged
\begin{equation}
\bar{s}_{\mu\si} =
\frac1{1 + 2 l_\mu}
\sum\limits_\nu s_{\nu\si} \delta_{l_\mu l_\nu} \delta_{k_\mu k_\nu}
\end{equation}
before the insertion into Eq.~(\ref{eq:F2}).

It is noteworthy that our model has the integrals of only two types:
either with all functions from the minimal set (all-minimal) as in Eq.~(\ref{eq:F0})
or with all but one from the minimal and one from the polarization set
(minimal-polarization one-electron and 3-minimal-1-polarization two-electron
integrals) as in Eq.~(\ref{eq:F1}).

The energy expression~(\ref{eq:E}) is minimized
under the orthogonality constraints~(\ref{eq:Corth})
to get the ground-state solution of the electronic structure problem,
the stationary conditions can be given in the form
of self-consistent-field (SCF) one-electron equations
\begin{equation}
\sum\limits_\nu \left(
\mathcal{F}_{\mu\nu\si} - \delta_{\mu\nu} \ep_{i\si}
\right) C_{\nu i\si} = 0
\end{equation}
with the effective Hamiltonian matrix derived as
\begin{equation}
\label{eq:FD}
\mathcal{F}_{\mu\nu\si} =
\partial E /\partial D_{\mu\nu\si} ,
\end{equation}
the explicit expression for the latter is lengthy but straightforward.
To compute a one-electron property $V$ such as dipole moment or
molecular electrostatic potential in a consistent way,
the energy derivative formalism should be used leading to
\begin{equation}
V =
  \sum\limits_{\mu\nu\si} V_{\mu\nu} D_{\mu\nu\si}
+ \sum\limits_{\al\nu} V_{\al\nu} D_{\al\nu}
\end{equation}
with the minimal-polarization block of the effective density matrix
given by the derivative
\begin{equation}
D_{\al\nu} = \partial E /\partial H_{\al\nu},
\end{equation}
its explicit form is again lengthy but straightforward.

We have written a computer code that solves these SCF equations
to get the energy and its first and second derivatives with respect to
the atomic coordinates and the applied uniform electric field,
the derivatives being computed in a fully analytic way.

\subsection{Molecular integrals}
\label{sec:MI}

The one- and two-electron integrals in the energy expression~(\ref{eq:E})
refer to the orthogonalized atom-centered basis functions~(\ref{eq:phi})
and as such are nontrivial functions
of (in general) all atomic coordinates of the molecular system.
They could have been computed from the first principles using
a three-level extension of our zero-bond-dipole orthogonalization scheme~\cite{L11}
applied to a fixed all-electron atomic basis --
the core set is orthogonalized first, the minimal valence set
is orthogonalized to the core and then within itself, and the polarization
set is orthogonalized to the core and valence and at last within itself.
Such first-principles integrals, however, can give, at best,
only a rough approximation to the Hartree-Fock molecular energy
within a polarized basis,
but we are aiming at a \textit{parametrizable} model that can yield
accurate molecular properties with the electron correlation included
in the spirit of Kohn-Sham density-functional theory.
Thus our model uses parametrized explicit analytical formulas
for the molecular integrals giving values that differ somewhat
from the first-principles ones.

The two-electron integrals are split into a sum
\begin{equation}
R_{\mu\nu\mu'\nu'} =
R^{(0)}_{\mu\nu\mu'\nu'} +
R^{(6)}_{\mu\nu\mu'\nu'}
\end{equation}
of "electrostatics" and "dispersion" parts,
the former quickly (exponentially) reaching the asymptotics
of Coulomb interaction of the two ($\mu\nu$ and $\mu'\nu'$) charge distributions
and the latter having a characteristic $r^{-6}$ asymptotic tail.
For the integrals with one polarization function
\begin{equation}
R_{\al\nu\mu'\nu'} =
R^{(0)}_{\al\nu\mu'\nu'}
\end{equation}
only the electrostatic part is used.
The electron correlation can already be modeled by decreasing the magnitude
of the two-electron integrals.

The one-electron integrals~(\ref{eq:H1})
have long-range Coulomb terms $V_{\mu\nu k'}$
and it is much easier to work with their short-range analogs
\begin{equation}
\bar{F}_{\mu\nu} = H_{\mu\nu} +
 \sum\limits_{\mu'} \left(
R^{(0)}_{\mu\nu\mu'\mu'} - \frct12 R^{(0)}_{\mu\mu'\mu'\nu}
\right) p_{\mu'} ,
\end{equation}
and in the same way for $\bar{F}_{\al\nu}$,
where a diagonal \textit{promolecule} density matrix
with constant spherically-symmetric atomic populations
\begin{equation}
p_\mu = p_{l_\mu k_\mu}
\end{equation}
is used to add the promolecule Coulomb and exchange terms,
so that they fully neutralize the core charges
\begin{equation}
\sum\limits_l p_{lk} = q_k .
\end{equation}
It is trivial to rewrite the energy expression
in terms of $\bar{F}$ and $R$ integrals,
working with the short-range integrals not only simplifies the parametrization
but is also of great advantage for large-scale calculations.
Moreover, some two-electron terms can be accounted for (on the average)
only within the $\bar{F}$ integrals and neglected in the $R$ integrals,
as we do in the following.
The (promolecule) one-electron integrals $\bar{F}_{\mu\nu}$
are of two kinds: one-center if $k_\mu = k_\nu$,
and two-center if $k_\mu \neq k_\nu$, but they should also depend
on the positions of the other atomic centers $k\neq k_\mu, k_\nu$.
In our model, we use an additive scheme
\begin{equation}
\bar{F}_{\mu\nu} =
\bar{F}_{\mu\nu,0} +
\sum\limits_{k\neq k_\mu, k_\nu}
 \bar{F}_{\mu\nu,k}
\end{equation}
where the first leading term is either one- or two-center and the sum runs over
the two- or three-center corrections, respectively.
The leading one-center integrals are atomic constants
\begin{equation}
\bar{F}_{\mu\nu,0}
\overset{k_\mu = k_\nu}{=}
\delta_{m_\mu m_\nu} \delta_{l_\mu l_\nu}
 F_{l_\mu k_\mu},
\end{equation}
\begin{equation}
\bar{F}_{\al\nu,0}
\overset{k_\al = k_\nu}{=}
0.
\end{equation}
The leading two-center integrals
\begin{equation}
\label{eq:Fmn}
\bar{F}_{\mu\nu,0}
\overset{k_\mu \neq k_\nu}{=}
\sum\limits_m
A_{mm_\mu}^{l_\mu}(\vec{z}_{\mu\nu})
A_{mm_\nu}^{l_\nu}(\vec{z}_{\mu\nu})
F_{ml_\mu l_\nu k_\mu k_\nu}(r_{\mu\nu})
\end{equation}
as well as $\bar{F}_{\al\nu,0}$
can be reduced to functions of interatomic distances
and the well-known transformation matrices
$A_{mm'}^l(\vec{z})$ of spherical harmonics upon spatial rotation,
\begin{equation}
\vec{z}_{\mu\nu} = \vec{r}_{\mu\nu}/r_{\mu\nu}, \qquad
\vec{r}_{\mu\nu} = \vec{r}_{k_\mu} - \vec{r}_{k_\nu}, \qquad
r_{\mu\nu} = |\vec{r}_{\mu\nu}|.
\end{equation}
The two-center corrections to the one-center integrals
\begin{equation}
\label{eq:Vmnk}
\bar{F}_{\mu\nu,k}
\overset{k_\mu = k_\nu \neq k}{=}
\sum\limits_{ml}
A_{mm_\mu}^{l_\mu}(\vec{z}_{\mu k})
A_{mm_\nu}^{l_\nu}(\vec{z}_{\mu k})
B_{0mm}^{l\:l_\mu l_\nu}
V_{ll_\mu l_\nu k_\mu k}(r_{\mu k})
\end{equation}
as well as $\bar{F}_{\al\nu,k}$
are done in the same way, making further use of the triple products
$B_{mm'm''}^{l\;\;l'\;\;l''}$ of spherical harmonics.
The most complicated are the three-center corrections $\bar{F}_{\mu\nu,k}$
with $k_\mu \neq k_\nu \neq k$ that can be exactly reduced only
down to nontrivial functions of three variables --
we have experimented with the triple series in prolate spheroidal coordinates
which can be made very accurate, but in the end we have chosen
the (approximate) factorization of the three-center terms
\begin{equation}
\label{eq:SSV}
\bar{F}_{\mu\nu,k}
\overset{k_\mu \neq k_\nu \neq k}{=}
\sum\limits_\la \delta_{k_\la k}
S_{\mu\la} S_{\nu\la} f_\la
\end{equation}
into the products of two-center terms
\begin{equation}
\label{eq:Smk}
S_{\mu\la}
\overset{k_\mu \neq k_\la}{=}
\sum\limits_m
A_{mm_\mu}^{l_\mu}(\vec{z}_{\mu\la})
A_{mm_\la}^{l_\la}(\vec{z}_{\mu\la})
S_{ml_\mu l_\la k_\mu k_\la}(r_{\mu\la})
\end{equation}
where $\la$ labels the functions of some expansion basis on center $k$
in which the underlying operator is diagonal with eigenvalues $f_\la$,
and $S_{\mu\la}$ are the overlap integrals.
The three-center terms with one polarization function
have minor effect and we set them to zero,
\begin{equation}
\bar{F}_{\al\nu,k}
\overset{k_\al \neq k_\nu \neq k}{=}
0.
\end{equation}

Of the two-electron integrals, only the one- and two-center ones
are of fundamental importance and are included in our model,
the rest are generally small and are set to zero.
The one-center integrals
\begin{equation}
R^{(0)}_{\mu\nu\mu'\nu'}
\overset{k_\mu = k_\nu = k_{\mu'} = k_{\nu'}}{=}
R^{(0)}_{\mu\nu\mu'\nu',0} +
\sum\limits_{k\neq k_\mu}
R^{(0)}_{\mu\nu\mu'\nu',k}
\end{equation}
have the leading term of atomic constants
\begin{equation}
\label{eq:glmnmn}
R^{(0)}_{\mu\nu\mu'\nu',0}
\overset{k_\mu = k_\nu = k_{\mu'} = k_{\nu'}}{=}
\sum\limits_{ml}
B_{mm_\mu m_\nu}^{l\;\; l_\mu \;\; l_\nu}
B_{mm_{\mu'} m_{\nu'}}^{l\;\; l_{\mu'} \;\; l_{\nu'}}
G_{ll_\mu l_\nu l_{\mu'} l_{\nu'}k_\mu}
\end{equation}
and additive corrections for the effect of surrounding atoms
\begin{equation}
\label{eq:Rmmk}
R^{(0)}_{\mu\nu\mu'\nu',k}
\overset{k_\mu = k_\nu = k_{\mu'} = k_{\nu'} \neq k}{=}
\delta_{m_\mu m_\nu} \delta_{m_{\mu'} m_{\nu'}}
\delta_{l_\mu l_\nu} \delta_{l_{\mu'} l_{\nu'}}
V_{l_\mu l_{\mu'} k_\mu k}(r_{\mu k}),
\end{equation}
for the latter we have also tested the general expression
\begin{equation}
R^{(0)}_{\mu\nu\mu'\nu',k}
\overset{k_\mu = k_\nu = k_{\mu'} = k_{\nu'} \neq k}{=}
\sum\limits_{mm'll'}
B_{mm_\mu m_\nu}^{l\;\; l_\mu \;\; l_\nu}
B_{m'm_{\mu'} m_{\nu'}}^{l'\;\; l_{\mu'} \;\; l_{\nu'}}
B_{0\: mm'}^{l''l\;\; l'}
V_{l''ll'l_\mu l_\nu l_{\mu'} l_{\nu'}k_\mu k}(r_{\mu k})
\end{equation}
but found only the terms with $l''=l=l'=0$ to be important,
we also find it safe to set
\begin{equation}
R^{(0)}_{\al\nu\mu'\nu'}
\overset{k_\al = k_\nu = k_{\mu'} = k_{\nu'}}{=}
0.
\end{equation}
Of the two-center two-electron integrals,
only the long-range ones
\begin{eqnarray}
\label{eq:Rmnmn}
R^{(0)}_{\mu\nu\mu'\nu'} &&
\overset{k_\mu = k_\nu \neq k_{\mu'} = k_{\nu'}}{=}
\sum\limits_{\bar{m}\bar{\bar{m}} \bar{m}'\bar{\bar{m}}' mll'}
A_{     \bar{m} m_\mu}^{l_\mu}(\vec{z}_{\mu\mu'})
A_{\bar{\bar{m}}m_\nu}^{l_\nu}(\vec{z}_{\mu\mu'})
A_{     \bar{m}' m_{\mu'}}^{l_{\mu'}}(\vec{z}_{\mu\mu'})
A_{\bar{\bar{m}}'m_{\nu'}}^{l_{\nu'}}(\vec{z}_{\mu\mu'})
\times
\nonumber \\
&& \times
B_{m\bar{m}\bar{\bar{m}}}^{l\;\;l_\mu l_\nu}
B_{m\bar{m}'\bar{\bar{m}}'}^{l'\;l_{\mu'} l_{\nu'}}
G_{mll'l_\mu l_\nu l_{\mu'} l_{\nu'} k_\mu k_{\mu'}}(r_{\mu\mu'}),
\end{eqnarray}
and in the same way $R^{(0)}_{\al\nu\mu'\nu'}$,
are included in the model, we studied the effect of other two-center terms
and found it safe to set them to zero.
We should stress that it is thanks to the special properties of the underlying
zero-bond-dipole~\cite{L11} orthogonalization of the basis functions
that all two-electron integrals involving two-center product charge distributions
$\phi_\mu(\vec{r}) \phi_\nu(\vec{r})$, $k_\mu \neq k_\nu$, 
are small and can be neglected.
The two-electron dispersion-model integrals
are naturally limited to the isotropic two-center terms
\begin{equation}
\label{eq:R6mnmn}
R^{(6)}_{\mu\nu\mu'\nu'}
\overset{k_\mu = k_\nu \neq k_{\mu'} = k_{\nu'}}{=}
\delta_{\mu \nu} \delta_{\mu' \nu'}
G^{(6)}_{l_\mu l_{\mu'} k_\mu k_{\mu'}}(r_{\mu\mu'}) .
\end{equation}
The zeroth-order eigenvalues in Eq.~(\ref{eq:xam})
can be optimized as atomic parameters,
but we find it enough to set
\begin{equation}
\epv_\mu = \bar{F}_{\mu\mu}, \qquad
\epv_\al = 1.
\end{equation}

Now that all molecular integrals in our model are defined in terms
of constants and functions of one variable $r$,
parametrized formulas for the latter are needed.
The short-range functions of Eqs.~(\ref{eq:Fmn}),
(\ref{eq:Vmnk}), (\ref{eq:Smk}), and (\ref{eq:Rmmk})
can be fitted to a high accuracy by the general expansion
\begin{equation}
\label{eq:fr}
f(r;a,c_0,\dots,c_n) =
\exp(-ar)\sum\limits_{\ka=0}^n (ar)^\ka c_\ka
\end{equation}
if enough terms are taken,
the long-range functions of Eq.~(\ref{eq:Rmnmn})
can be done likewise with one long-range term added
\begin{equation}
\label{eq:gl}
g_l(r;q,a,c_0,\dots,c_n) =
q r^{-1-l} u_{2l}(ar)
+ a^{1+l} \exp(-ar)\sum\limits_{\ka=0}^n (ar)^{l+\ka} c_\ka
\end{equation}
with
\begin{equation}
u_n(r) = 1 - \exp(-r) \sum\limits_{m=0}^n \frac{r^m}{m!} ,
\end{equation}
and the dispersion tail corrections~(\ref{eq:R6mnmn})
can easily be added in the form
\begin{equation}
g^{(6)}(r;a,c) =
c r^{-6} u_9(ar) .
\end{equation}
We have experimented with these general expansions within our
electronic structure model
(working with more than 20000 atomic and atom-pair parameters
for a set of 7 chemical elements!), but found later that much more
compact expressions using sum rules with mostly atomic parameters
work quite well and yield more meaningful optimized parameter values.
The short-range terms can be given either
by a single exponential ($n=0$) term of Eq.~(\ref{eq:fr})
or by a three-parameter hyperbolic-secant function
\begin{equation}
f_\textrm{o}(r;a,b,c) =
c \exp (-ar+b) \left(1 + \exp(-2ar+2b)\vphantom{^1}\right)^{-1}
\end{equation}
with $b>0$. For the long-range terms, the leading term of Eq.~(\ref{eq:gl})
is enough for all multipole-multipole and charge-multipole interactions
and only a single exponential term needs to be added
for the charge-charge interactions.

The sum-rule formulas used in this work are as follows.
In Eq.~(\ref{eq:Fmn}) we have
\begin{equation}
\label{eq:Fmnr}
F_{m l_\mu l_\nu k_\mu k_\nu}(r) =
f\left(r;
\sqrt{a_\mu a_\nu}\; a_{m l_\mu l_\nu \xi_\mu \xi_\nu},
c_\mu c_\nu c_{m l_\mu l_\nu \xi_\mu \xi_\nu}
\right)
\end{equation}
with the shorthand notation for atomic parameters
$a_\mu \equiv a_{l_\mu k_\mu}$, $c_\mu \equiv c_{l_\mu k_\mu}$,
and $\xi_\mu$ being atom-group labels -- in particular, $\xi_\mu = 1$
if $\mu$ is on H, $\xi_\mu = 2$ if $\mu$ is on a second-row atom Li--F,
and $\xi_\mu = 3$ for $\mu$ on Na--Cl.
Thus we work with atomic parameters
and diatomic atom-group parameters $a_{m l_\mu l_\nu \xi_\mu \xi_\nu}$,
$c_{m l_\mu l_\nu \xi_\mu \xi_\nu}$ in Eq.~(\ref{eq:Fmnr}),
and this is also done in the formulas below.
For minimal-polarization analog of Eq.~(\ref{eq:Fmn})
we choose
\begin{equation}
\label{eq:Fanr}
F_{m l_\al l_\nu k_\al k_\nu}(r) =
f_\textrm{o}\left(r;
\frac{2 a_\al a_\nu}{a_\al + a_\nu}
a_{m l_\al l_\nu \xi_\al \xi_\nu },
b_{\xi_\mu \xi_\nu},
c_\al c_\nu c_{m l_\al l_\nu \xi_\al \xi_\nu }
\right)
\end{equation}
and it should be understood that, here and below,
each integral class has its own set
of parameters, for example $a_\nu$ in Eq.~(\ref{eq:Fmnr})
is not the same as $a_\nu$ in Eq.~(\ref{eq:Fanr}) --
the same notation is used to save space.
In Eq.~(\ref{eq:Vmnk}) we choose the form
\begin{equation}
V_{ll_\mu l_\nu k_\mu k}(r) =
f_\textrm{o}\left(r;
\frac{2 (a_\mu + a_\nu) a_k}{a_\mu + a_\nu + 2a_k}
a_{l l_\mu l_\nu \xi_\mu \xi_k},
b_{\xi_\mu \xi_k},
\frct12 (c_\mu + c_\nu) c_k c_{l l_\mu l_\nu \xi_\mu \xi_k}
\right)
\end{equation}
for both minimal-minimal and minimal-polarization integrals.
In Eq.~(\ref{eq:Smk}) we have
\begin{equation}
S_{m l_\mu l_\la k_\mu k_\la}(r) =
f_\textrm{o}\left(r;
\frac{2 a_\mu a_\la}{a_\mu + a_\la},
b_{\xi_\mu \xi_\la},
c_\mu c_\la c_{m l_\mu l_\la \xi_\mu \xi_\la }
\right).
\end{equation}
In Eq.~(\ref{eq:Rmmk}) we have
\begin{equation}
V_{l_\mu l_{\mu'} k_\mu k}(r) =
f_\textrm{o}\left(r;
\frac{2 (a_\mu + a_{\mu'}) a_k}{a_\mu + a_{\mu'} + 2a_k},
b_{\xi_\mu \xi_k},
\frct12 (c_\mu + c_{\mu'}) c_k c_{l_\mu l_{\mu'} \xi_\mu \xi_k}
\right).
\end{equation}
In Eq.~(\ref{eq:Rmnmn}) the form
\begin{equation}
\label{eq:gmll}
G_{mll'l_\mu l_\nu l_{\mu'} l_{\nu'} k_\mu k_{\mu'}}(r) =
g_{l+l'}\left(r;
q_{l l_\mu l_\nu} q_{l' l_{\mu'} l_{\nu'}} q_{mll'},
\frac{2(a_\mu + a_\nu)(a_{\mu'} + a_{\nu'})}{a_\mu + a_\nu + a_{\mu'} + a_{\nu'}},
-\frct12 \delta_{0l} \delta_{0l'}
\right)
\end{equation}
is used for both 4-minimal and 1-polarization-3-minimal two-electron integrals,
with the atomic multipoles $q_{l l_\mu l_\nu}$ and
the fundamental constants of multipole-multipole interaction $q_{mll'}$,
in the charge-charge 4-minimal case one exponential term is added.
In Eq.~(\ref{eq:R6mnmn}) we have
\begin{equation}
\label{eq:g6}
G^{(6)}_{l_\mu l_{\mu'} k_\mu k_{\mu'}}(r) =
g^{(6)}\left(r;
\frac{2 a_\mu a_{\mu'}}{a_\mu + a_{\mu'}},
c_\mu c_{\mu'} \right).
\end{equation}
As can be seen, the atomic $a$-parameters in all these formulas
play the role of radial scale factors, the atomic $c$-parameters
are multiplicative prefactors, and the atom-group $b$-parameters
control the shape of functions at shorter distances $r$.

\subsection{Parametrization procedure}
\label{sec:P}

We parametrize our model by minimizing the function
\begin{equation}
\label{eq:Esum}
\mathcal{E} =
\mathcal{E}_E +
\mathcal{E}_F +
\mathcal{E}_D +
\mathcal{E}_g +
\mathcal{E}_H +
\mathcal{E}_d +
\mathcal{E}_a +
\mathcal{E}_q +
\mathcal{E}_U +
\mathcal{E}_6 +
\mathcal{E}_c
\end{equation}
that measures the deviation of molecular properties
predicted by the model from the reference values
computed by the higher-level theory
on a set of molecules (molecular geometries).
The energy term
\begin{equation}
\label{eq:EE}
\mathcal{E}_E =
\sum\limits_n w^E_n
\left( \sum\limits_m c_{mn} (\tilde{E}_m - E_m)
\right)^2 ,
\end{equation}
with $m=1,\dots,M$ labeling each molecule,
can be used not only to fit each predicted energy $\tilde{E}_m$
to the reference value $E_m$ 
in the trivial case $c_{mn}=\delta_{mn}$,
but also for giving higher weights $w^E_n$ to some chemically-meaningful
energy differences, such as conformational energy changes,
reaction energies and activation barriers.

The effective one-electron Hamiltonian terms
\begin{equation}
\label{eq:EF}
\mathcal{E}_F =
\sum\limits_m w^F_m \tr\{(\tilde{\vec{F}}_m - \vec{F}_m)^2 \}
\end{equation}
and the density-matrix terms
\begin{equation}
\label{eq:ED}
\mathcal{E}_D =
\sum\limits_m w^D_m \tr\{(\tilde{\vec{D}}_m - \vec{D}_m)^2 \}
\end{equation}
for each molecule take $\tilde{\vec{F}}_m$ from Eq.~(\ref{eq:FD})
and $\vec{F}_m$ from the theory of Section~\ref{sec:FD} below,
the density matrices $\tilde{\vec{D}}_m$ and $\vec{D}_m$
come from the diagonalization of $\tilde{\vec{F}}_m$ and $\vec{F}_m$.
One may put all $w^D_m=0$ and work with only $w^F_m \neq 0$,
assuming that an accurate $\vec{F}$-matrix should also yield
an accurate $\vec{D}$-matrix,
but in practice we have not-so-small errors in $\vec{F}$-matrices
and find it a better compromise to balance between $\vec{F}$ and $\vec{D}$ terms,
thus giving more importance to the occupied-occupied and occupied-virtual blocks
of $\vec{F}$.

The energy first $\vec{g}_m$ and second $\vec{H}_m$ derivatives with respect
to atomic coordinates for each molecule $m$ are gathered into the terms
\begin{equation}
\label{eq:Eg}
\mathcal{E}_g =
\sum\limits_m w^g_m
(\tilde{\vec{g}}_m - \vec{g}_m)^\te \vec{W}^2_m (\tilde{\vec{g}}_m - \vec{g}_m) ,
\end{equation}
\begin{equation}
\label{eq:Eh}
\mathcal{E}_H =
\sum\limits_m w^H_m \tr\left\{
\left((\tilde{\vec{H}}_m - \vec{H}_m) \vec{W}_m \right)^2 \right\} ,
\end{equation}
where the weight matrix
\begin{equation}
\label{eq:Wh}
\vec{W} = \left(\vec{H}^2 + h_0^2 \vec{1}\right)^{-1/2}
\end{equation}
is a well-behaved positive-definite replacement for the inverse $\vec{H}^{-1}$.
This weighting scheme emphasizes the weak modes and, in our experience,
it drives the model towards a balanced reproduction of potential energy surfaces.
We set $h_0=2^{-10}$ in this work, close to the lowest eigenvalue
of $\vec{H}$ for the water dimer (H$_2$O)$_2$ used as a prototype.
For reference geometries at stationary points ($\vec{g}_m=0$),
as is normally the case,
Eq.~(\ref{eq:Eg}) can be seen as a weighted sum of squared distances
between the stationary points estimated by the model and the reference,
whereas Eq.~(\ref{eq:Eh}) is a weighted sum of squared dimensionless
relative deviations of force constant values.

Electric dipole moments and polarizabilities
\begin{equation}
\label{eq:Ed}
\mathcal{E}_d =
\sum\limits_m w^d_m \left|\tilde{\vec{d}}_m - \vec{d}_m\right|^2,
\end{equation}
\begin{equation}
\label{eq:Ea}
\mathcal{E}_a =
\sum\limits_m w^a_m \tr\{(\tilde{\vec{a}}_m - \vec{a}_m)^2\}
\end{equation}
as well as atomic polar tensors
\begin{equation}
\label{eq:Eq}
\mathcal{E}_q =
\sum\limits_m w^q_m
 \tr\{(\tilde{\vec{q}}_m - \vec{q}_m)^\te (\tilde{\vec{q}}_m - \vec{q}_m)\}
\end{equation}
are included into the optimization in a straightforward way,
as are the molecular electrostatic potentials
\begin{equation}
\label{eq:EU}
\mathcal{E}_U =
\sum\limits_m w^U_m
\sum\limits_\ka \left(\tilde{U}(\vec{r}_\ka) - U(\vec{r}_\ka)\right)^2 s_\ka
\end{equation}
on the grids of points $\vec{r}_\ka$ at discretized surfaces
with elements $s_\ka$ (see also Section~\ref{sec:SV}).

Molecular dispersion coefficients (discussed in Section~\ref{sec:C6} below)
are treated in the form of square roots
\begin{equation}
\label{eq:E6}
\mathcal{E}_6 =
\sum\limits_m w^{(6)}_m \left(\tilde{c}_{6,m}^{1/2} - c_{6,m}^{1/2}\right)^2
\end{equation}
as these have a more natural system-size scaling.

The last (optional) term is a sum of soft constraints
\begin{equation}
\label{eq:Ec}
\mathcal{E}_c =
\sum\limits_\ka w^c_\ka
s \left( \sum\limits_p c_{p\ka} x_p ,
 \bar{c}_\ka, \bar{\bar{c}}_\ka
\right),
\end{equation}
\begin{equation}
s(x,a,b) = \left\{
\begin{array}{rll}
(x-a)^2,& & x < a \\
 0     ,& & a < x < b \\
(x-b)^2,& & b < x
\end{array}
\right.
\end{equation}
on linear combinations of optimization parameters $\{x_p\}$.

\subsection{Effective Hartree-Fock-like
Hamiltonian from correlated wavefunction theory}
\label{sec:FD}

Given the one-electron nonidempotent density matrices $\mathcal{D}_{\mu\nu\si}$
from a correlated wavefunction calculation, with $\mu, \nu$ labeling
(in this Section only) the functions of an extended atomic basis,
we set up our Hartree-Fock-like self-consistent field procedure
based on the minimization of the energy expression
\begin{eqnarray}
\label{eq:Ew}
E^{(w)} &=&
 \sum\limits_{\mu\nu\si} D_{\mu\nu\si}^{(w)} H_{\mu\nu}
+ \frct12\!\! \sum\limits_{\mu\nu\si\mu'\nu'\si'}
 D_{\mu\nu\si}^{(w)} R_{\mu\nu\si\mu'\nu'\si'} D_{\mu'\nu'\si'}^{(w)}
\nonumber
\\
&+&
 \frac{w}2 \sum\limits_{\mu\nu\mu'\nu'\si}
 \left( D_{\mu\nu\si}^{(w)} - \mathcal{D}_{\mu\nu\si} \right)
 R_{\mu\nu\mu'\nu'}
 \left( D_{\mu'\nu'\si}^{(w)} - \mathcal{D}_{\mu'\nu'\si} \right)
\end{eqnarray}
with respect to the idempotent density matrices $D_{\mu\nu\si}$.
Eq.~(\ref{eq:Ew}) is a sum of the Hartree-Fock energy
and a quadratic density penalty function weighted by $w$,
the associated effective Hamiltonian matrix
\begin{equation}
F_{\mu\nu\si}^{(w)} = H_{\mu\nu}
+ \sum\limits_{\mu'\nu'\si'} R_{\mu\nu\si\mu'\nu'\si'} D_{\mu'\nu'\si'}^{(w)}
+ w \sum\limits_{\mu'\nu'} R_{\mu\nu\mu'\nu'}
 \left( D_{\mu'\nu'\si}^{(w)} - \mathcal{D}_{\mu'\nu'\si} \right)
\end{equation}
is a sum of the Fock matrix and a local spin-dependent correlation potential matrix,
the latter arises as the scaled Coulomb potential of the difference spin density.
If a complete basis were used, the limit $w\to \infty$, if exists,
should yield the exact effective Hamiltonian
of Sec.~IIB of Kohn and Sham's work~\cite{KS65}.
In practice, we work with a finite incomplete basis
and a reasonable finite value of $w$ should be chosen.
Our procedure can be seen as a nonlocal-exchange-local-correlation analog
of the local-exchange-correlation procedure of Zhao, Morrison, and Parr~\cite{ZMP94}.

\subsection{Molecular dispersion coefficients}
\label{sec:C6}

For two molecules at a large distance $r$ between their centers, the leading
asymptotic term of the dispersion part of the intermolecular potential
is $c_6 r^{-6}$ with $c_6$ being a function of the relative orientation.
The orientational average of $c_6$ is a bimolecular constant
that can be most easily computed within MP2 as
\begin{equation}
\label{eq:c6}
c_6 = \frct23
 \sum\limits_{ai\si}^{\textrm{mol.} 1}
 \sum\limits_{a'i'\si'}^{\textrm{mol.} 2}
\frac{
\left|\left<\phi_{a\si}\right|\vec{r}\left|\phi_{i\si} \right>\right|^2
\left|\left<\phi_{a'\si'}\right|\vec{r}\left|\phi_{i'\si'} \right>\right|^2
}{\ep_{i\si} - \ep_{a\si} + \ep_{i'\si'} - \ep_{a'\si'}}
\end{equation}
from the dipole moment integrals over one-electron wavefunctions $\phi$
and the one-electron energies $\ep$,
$i$'s label the occupied and $a$'s the virtual states
each localized on one of the molecules.
Much more complicated expressions can be derived for higher-order
correlation methods, but we will use Eq.~(\ref{eq:c6}) in this work
as it gives accurate enough values for our purpose.
We take only the case of two identical molecules and such $c_6$
becomes a molecular constant of interest, moreover, we have seen
that a heteromolecular $c_6$ is quite close
to the geometric mean of two molecular constants.
Within our parametrizable model the value to put into Eq.~(\ref{eq:E6})
is simply
\begin{equation}
\tilde{c}_6^{1/2} =
\sum\limits_{\mu\si} D_{\mu\mu\si} c_\mu
\end{equation}
with $c_\mu$ as in Eq.~(\ref{eq:g6}). By construction, Eq.~(\ref{eq:g6})
always gives isotropic $c_6$ values that follow the rule of geometric mean.

\subsection{Surfaces for sampling molecular electrostatic potentials}
\label{sec:SV}

A smooth surface is preferable for sampling the molecular electrostatic potential,
so we have tried the isodensity surface $\rho(\vec{r})=\rho_0$
and found that $\rho_0\approx10^{-6}$ should be used to enclose
most of the density as needed. While this may work well for the exact density,
the one calculated within a limited finite basis approximation
may not be accurate enough at such small values in the tail region,
indeed we have seen some rather weird shapes for molecules as simple as LiF.

In this work we set a new definition of molecular surface
$p(\vec{r})=p_0$ with
\begin{equation}
\label{eq:pr}
p(\vec{r}) =
 \int
 s(\vec{r}'-\vec{r})
 \rho_1(\vec{r}',\vec{r}'')
 s(\vec{r}''-\vec{r})
\de^3 \vec{r}'
\de^3 \vec{r}''
\end{equation}
in terms of the one-electron density matrix $\rho_1(\vec{r},\vec{r})$,
the density being its diagonal part $\rho(\vec{r})=\rho_1(\vec{r},\vec{r})$,
as inspired by the analysis of the exchange repulsion effects.
For the broadening function $s(\vec{r})$ localized around $\vec{r}=0$
we make the simplest choice
\begin{equation}
\label{eq:sr}
s(\vec{r}) = c \exp\left(a|\vec{r}|^2\right),
\end{equation}
in the limit $a\to \infty$ with $c=(a/\pi)^{3/2}$
Eqs.~(\ref{eq:pr}) and~(\ref{eq:sr}) yield $p(\vec{r})\to \rho(\vec{r})$.
With $c=1$, $a=1/16$, and $p_0=1/4$ we get good surfaces for the molecules
studied in this work.
The surface discretization is best done
with the spherical quadrature rules~\cite{L76,LL99}.

\section{Calculations}
\label{sec:calc}

The training set of 5581 molecules used in this work
covers a broad range of structures
built up from 15 elements H, Li--F, Na--Cl,
there are both energy-minimum and transition-state geometries
of neutral, cationic and anionic species
with closed-shell and (high-spin) open-shell electronic configurations
dominated by a single determinant.
We have spent a lot of time setting up this database guided by
our chemical intuition, finding meaningful structures
of (almost) all chemical compositions with up to three nonhydrogen
and any number of hydrogen atoms,
and less thoroughly for up to eight nonhydrogen atoms.
We tried to sample all kinds of chemical bonds --
from metallic through covalent to ionic, 
hydrogen bonded dimers and clusters
(for example, neutral, protonated, and deprotonated water clusters
with up to 6 O atoms) as well as weaker intermolecular complexes
are also carefully chosen.

The reference data were generated by the CCSD method~\cite{PB82}
with nonrelativistic Hamiltonian and
correlation-consistent atomic basis sets~\cite{L05}
of sizes shown in Table~\ref{tab:abcc}.
The \texttt{L1} set has one set of valence polarization functions,
the \texttt{L11} set is for a correlated treatment of (outermost) core shells,
and the \texttt{L1+1} set has diffuse functions
added as Rydberg shells --
this is the entry-level basis for quantitative CCSD calculations,
we would have preferred the next-level (\texttt{L2}, \texttt{L22}, \texttt{L2+1})
had we had enough computer power.
An archive of all molecular data files is available for download~\cite{epaps}.

The model is defined by the energy expression of Section~\ref{sec:E},
the molecular integrals of Section~\ref{sec:MI},
and the atomic basis set sizes given in Table~\ref{tab:abm}
with one radial function for each angular symmetry.
Besides the full model, 8 simplified models with some classes of integrals
set to zero are also analyzed.
All parameters of each model have been simultaneously optimized
on the full reference data set,
the weights assigned to each molecular property
and the values of each error term in Eq.~(\ref{eq:Esum}) at convergence
are given in Table~\ref{tab:err},
full details of the settings along with the optimized parameter values
can be found in the attached files~\cite{epaps}.
Our optimization algorithm computes numerically
all first derivatives of molecular properties with respect to the parameters
and uses linear searches along the optimal direction to find the nearest local
minimum, the choice of the starting guess is so far from straightforward
that it cannot be documented here.

It is not trivial to judge the quality of such a model by the net error terms,
but still some insight into the role of the whole classes
of parametrized molecular integrals can be gotten
as they are removed from the model in the order of increasing importance.
As expected, the dispersion terms~(\ref{eq:R6mnmn}) have the smallest impact
as shown by Model~2. The role of the polarization basis functions is dramatic --
Model~4 that has them all removed nearly doubles the most important error terms,
and about 2/3 of this effect come
from the one-electron integrals~(\ref{eq:Fmn}) and~(\ref{eq:Vmnk}),
Model~3 with only the two-electron integrals~(\ref{eq:Rmnmn})
over the polarization functions included
is the simplest one to properly account for the molecular dipole polarizability
but this does also improve the other molecular properties.
Starting from Model~4 we can study the role of the integrals over the minimal basis
that are new to our model.
The two-center corrections to one-center two-electron integrals~(\ref{eq:Rmmk})
are the least important, though Model~5 without them shows already
about 1.5 greater values of most error functions.
Further degradation is caused by zeroing out the multipole moments
in the two-center two-electron integrals~(\ref{eq:gmll}) as in Model~6
and, further on,
in the one-center two-electron integrals~(\ref{eq:glmnmn}) as in Model~7.
At this point, Model~7 is already too simplified to support the burden
of the first~(\ref{eq:Eg}) and second~(\ref{eq:Eh}) derivative terms
as they were before, so we need to relieve it by setting a larger value of $h_0$
in Model~7$'$ to get ready for the last strike.
A dramatic degradation is seen as the three-center corrections
to two-center one-electron integrals~(\ref{eq:SSV})
are zeroed out in Model~8, and yet again as the two-center corrections
to one-center one-electron integrals~(\ref{eq:Vmnk}) are dropped in Model~9.
With this overview,
it should be clear that these integral terms have
not only the quantitative but also the qualitative impact on the performance
of the model -- the simplified models may even fail to reproduce the existence
of some stationary point on molecular potential energy surfaces,
for example, Model~6 predicts the symmetry
of the hydrogen peroxide molecule H$_2$O$_2$
to be C$_\textrm{2h}$ instead of C$_2$,
and Model~8 gives the cyclobutane molecule C$_4$H$_8$
the D$_\textrm{4h}$ symmetry instead of the right D$_\textrm{2d}$
-- the reliable prediction of intra- and intermolecular conformations
is a challenge that only our full model seems to meet.

A binary executable code of our program for Linux/x86\_64 architecture
is made available~\cite{epaps} to the interested researchers worldwide
for further testing and some preliminary applications of the new model.

\section{Conclusions}

We are pleased to have developed
the new parametrizable electronic structure model
that both has a number of good formal properties
and performs well in numerical tests.
It evolved through a trial-and-error process
until it has reached a satisfactory level of maturity.
With this parameter set for 15 elements,
many interesting systems can already be studied,
we are going to see how it works
for the structure prediction of biomolecules
and the chemical processes in condensed phase.
If some weaknesses will be found,
the model can be reparametrized on a more representative training set.

Future work may include:
building a comprehensive database of prototype molecular structure
for further training and testing of the model,
its extension to heavier elements,
a reparametrization based on more accurate reference data
from CCSD or CCSD(T) calculations with a next-level basis,
the investigation of linear-scaling
techniques~\cite{S96,DMS97,AABBA04,LYY96,DM96,DS99} for solving the SCF equations
of our model to find a both accurate and fast algorithm
for studies of large systems.
A formalism to treat excited electronic states
dominated by single excitations from a single determinant
can also be worked out, either
within the configuration interaction with single substitutions (CIS)
or within the linear response theory.

\begin{acknowledgments}
It took us 5 (five) years of hard work to come up
with this new electronic structure model.
\end{acknowledgments}

\clearpage

\begin{center}
\squeezetable
\begin{table} 
\caption{Atomic basis set sizes for CCSD calculations.}
\label{tab:abcc}
\begin{ruledtabular}
\begin{tabular}{lllll}
Atom      & Set  & Core & Contracted & Primitive \\
\hline
H         & \texttt{L1}   &      & 2,1   & 8,4     \\
Li, Be    & \texttt{L11}  &      & 4,3,1 & 12,8,4  \\
B, C      & \texttt{L1}   & 1    & 3,2,1 & 12,8,4  \\
N, O, F   & \texttt{L1+1} & 1    & 4,3,1 & 14,10,4 \\
Na, Mg    & \texttt{L11}  & 1    & 5,4,2 & 18,13,8 \\
Al, Si, P & \texttt{L1}   & 2,1  & 4,3,1 & 18,13,5 \\
S, Cl     & \texttt{L1+1} & 2,1  & 5,4,1 & 20,15,5 \\
\end{tabular}\\
Number of radial functions for each angular quantum number is given.
\end{ruledtabular}
\end{table}
\end{center}

\clearpage

\begin{center}
\squeezetable
\begin{table} 
\caption{Atomic basis set sizes in the parametrized model.}
\label{tab:abm}
\begin{ruledtabular}
\begin{tabular}{lllll}
Atom             & $l_\mu^\textrm{max}$ & $l_\al^\textrm{max}$ & $l_\la^\textrm{max}$ \\
\hline
H                & 0  & 1 & 1 \\
Li, Be           & 1  & 1 & 2 \\
B, C, N, O, F    & 1  & 2 & 2 \\
Na, Mg           & 1  & 1 & 2 \\
Al, Si, P, S, Cl & 1  & 2 & 2 \\
\end{tabular}\\
Maximum angular quantum number
in the minimal $l_\mu^\textrm{max}$,
polarization $l_\al^\textrm{max}$,
and projection~(\ref{eq:SSV}) $l_\la^\textrm{max}$ basis.
\end{ruledtabular}
\end{table}
\end{center}

\clearpage

\begin{center}
\squeezetable
\begin{table} 
\caption{Errors in molecular properties from the parametrized models.}
\label{tab:err}
\begin{ruledtabular}
\begin{tabular}{lclrrrrrrrrrrr}
Integral term$^a$               && Eq.             & Model 1 & Model 2 & Model 3 & Model 4 & Model 5 & Model 6 & Model 7 & Model 7$'$& Model 8 & Model 9 \\
\multicolumn{2}{l}{$\bar{F}_{\mu\nu,k}^{(k_\mu = k_\nu \neq k)}$}
                                 &(\ref{eq:Vmnk})  & {\bf+}  & {\bf+}  & {\bf+}  & {\bf+}  & {\bf+}  & {\bf+}  & {\bf+}  & {\bf+}  & {\bf+}  & {\bf }  \\
\multicolumn{2}{l}{$\bar{F}_{\mu\nu,k}^{(k_\mu \neq k_\nu \neq k)}$}
                                 &(\ref{eq:SSV})   & {\bf+}  & {\bf+}  & {\bf+}  & {\bf+}  & {\bf+}  & {\bf+}  & {\bf+}  & {\bf+}  & {\bf }  & {\bf }  \\
\multicolumn{2}{l}{$G_{ll_\mu l_\nu l_{\mu'} l_{\nu'}k_\mu}^{(l>0,k_\mu = k_\nu = k_{\mu'} = k_{\nu'})}$}
                                 &(\ref{eq:glmnmn})& {\bf+}  & {\bf+}  & {\bf+}  & {\bf+}  & {\bf+}  & {\bf+}  & {\bf }  & {\bf }  & {\bf }  & {\bf }  \\
\multicolumn{2}{l}{$q_{l l_\mu l_\nu}^{(l > 0)}$}
                                 &(\ref{eq:gmll})  & {\bf+}  & {\bf+}  & {\bf+}  & {\bf+}  & {\bf+}  & {\bf }  & {\bf }  & {\bf }  & {\bf }  & {\bf }  \\
\multicolumn{2}{l}{$R_{\mu\nu\mu'\nu',k}^{(0)(k_\mu = k_\nu = k_{\mu'} = k_{\nu'} \neq k)}$}
                                 &(\ref{eq:Rmmk})  & {\bf+}  & {\bf+}  & {\bf+}  & {\bf+}  & {\bf }  & {\bf }  & {\bf }  & {\bf }  & {\bf }  & {\bf }  \\
\multicolumn{2}{l}{$\bar{F}_{\al\nu,0}^{(k_\al \neq k_\nu)}$}
                                 &(\ref{eq:Fmn})   & {\bf+}  & {\bf+}  & {\bf }  & {\bf }  & {\bf }  & {\bf }  & {\bf }  & {\bf }  & {\bf }  & {\bf }  \\
\multicolumn{2}{l}{$\bar{F}_{\al\nu,k}^{(k_\al = k_\nu \neq k)}$}
                                 &(\ref{eq:Vmnk})  & {\bf+}  & {\bf+}  & {\bf }  & {\bf }  & {\bf }  & {\bf }  & {\bf }  & {\bf }  & {\bf }  & {\bf }  \\
\multicolumn{2}{l}{$R_{\al\nu\mu'\nu',0}^{(0)(k_\al = k_\nu \neq k_{\mu'} = k_{\nu'})}$}
                                 &(\ref{eq:Rmnmn}) & {\bf+}  & {\bf+}  & {\bf+}  & {\bf }  & {\bf }  & {\bf }  & {\bf }  & {\bf }  & {\bf }  & {\bf }  \\
\multicolumn{2}{l}{$R^{(6)}_{\mu\nu\mu'\nu'}$}
                                 &(\ref{eq:R6mnmn})& {\bf+}  & {\bf }  & {\bf+}  & {\bf+}  & {\bf }  & {\bf }  & {\bf }  & {\bf }  & {\bf }  & {\bf }  \\
\multicolumn{2}{l}{Number of parameters$^b$}
                                               &                 &     720 &     690 &     524 &     513 &     434 &     406 &     378 &     378 &     264 &     185 \\
Error term$^c$ &\multicolumn{1}{c}{Eq.}        & $w$ $^d$        &         &         &         &         &         &         &         &         &         &         \\
$\mathcal{E}_{E_\textrm{total}}$ &(\ref{eq:EE})& $2^{13} $       &   975.8 &  1013.0 &  1718.5 &  2206.1 &  3307.7 &  5006.1 &  5940.6 &  4498.5 &  8176.7 & 12026.6 \\
$\mathcal{E}_{E_\textrm{react}}$ &(\ref{eq:EE})& $2^{22} $       &   228.1 &   255.1 &   497.8 &   562.6 &  1022.6 &  1459.8 &  1662.8 &   591.4 &  3010.9 &  4251.5 \\
$\mathcal{E}_F$                  &(\ref{eq:EF})& $2^{-2} $       &   998.3 &   989.6 &  1131.8 &  1190.0 &  1013.9 &  1107.5 &  1266.4 &   902.5 &  1684.5 &  1811.7 \\
$\mathcal{E}_D$                  &(\ref{eq:ED})& $1      $       &   267.2 &   283.1 &   522.7 &   552.3 &   675.7 &   771.5 &  1042.8 &   720.9 &  1318.9 &  1115.3 \\
$\mathcal{E}_g$                  &(\ref{eq:Eg})& $2^5    $       &  3194.3 &  3255.2 &  5067.6 &  5978.6 &  9712.6 & 12905.1 & 13887.4 &  3432.9 &  5439.0 &  8111.3 \\
$\mathcal{E}_H$                  &(\ref{eq:Eh})& $1      $       &  1961.9 &  1987.2 &  3398.4 &  3957.8 &  6313.1 &  7774.1 &  9329.4 &  1872.7 &  2990.5 &  4356.6 \\
$\mathcal{E}_d$                  &(\ref{eq:Ed})& $2^{-3} $       &     7.5 &     7.6 &    11.7 &    22.6 &    31.5 &    56.9 &    63.6 &    47.2 &    86.7 &   119.7 \\
$\mathcal{E}_a$                  &(\ref{eq:Ea})& $2^{-10}$       &    62.6 &    58.0 &    70.0 &    --   &    --   &    --   &    --   &    --   &    --   &    --   \\
$\mathcal{E}_q$                  &(\ref{eq:Eq})& $1      $       &   196.1 &   199.9 &   289.3 &   477.6 &   615.7 &   661.4 &   782.9 &   669.1 &  1315.5 &  1258.3 \\
$\mathcal{E}_U$                  &(\ref{eq:EU})& $2^3    $       &   324.3 &   324.5 &   332.2 &   331.6 &   305.9 &   262.3 &   255.6 &   278.5 &   263.5 &   232.2 \\
$\mathcal{E}_6$                  &(\ref{eq:E6})& $2^{-3} $       &    85.4 &    --   &   200.4 &   284.2 &    --   &    --   &    --   &    --   &    --   &    --   \\
                                 &(\ref{eq:Wh})& $h_0$ $^e$      &$2^{-10}$&$2^{-10}$&$2^{-10}$&$2^{-10}$&$2^{-10}$&$2^{-10}$&$2^{-10}$& $2^{-5}$& $2^{-5}$& $2^{-5}$\\
\end{tabular}\\
\begin{flushleft}
$^a$The integral terms of the Eqs. given are included in the model
if marked by the "{\bf+}" sign, otherwise set to zero.\\
$^b$The number of independent parameters optimized for each model.\\
$^c$The error terms of the Eqs. given with $^d$the assigned weights
and $^e$the $h_0$ values of Eq.~(\ref{eq:Wh}) are listed (in au)
for each model if included in the optimization,
otherwise assigned zero weight and marked by "--".
\end{flushleft}
\end{ruledtabular}
\end{table}
\end{center}

\clearpage

\noindent
{\large JCP Manuscript \textbf{A11.06.0280}}

\noindent
Title: \textbf{A new parametrizable model of molecular electronic structure}

\noindent
Author: \textbf{Dimitri N. Laikov}

\begin{center}
\textbf{Discussion of Reviewers' Comments and Changes made to the Maniscript}
\end{center}

Many thanks to the Reviewers for the careful reading of the Manuscript
and the valuable comments and suggestions.
I have an impression that my work has excited a lot of interest --
the Reviewers seem to be eager to see more details on the performance
of the new model, so they both ask for more tests of the model's accuracy
to be done and reported right now.
I strongly believe that such extended tests would make up a very good report
publishable separately, not to say that I have limited time left would I have wished
to do it now, moreover, I would be delighted to see such testing
done by a third party.
I find a good way to resolve this problem -- the Supplementary Information
will now include the binary executable file (single-threaded only)
of my computer code for the most common machine architecture (Linux/x86\_64)
that can be used by anyone in the world to run molecular calculations
with the new model -- geometry optimizations of stable and transition states,
harmonic vibrational analyses, intrinsic reaction coordinates --
sample input files are also included and an input description is added
to the README file.
On the other hand, the Reviewers found no flaw in the most fundamental part
of the work -- the mathematics of the new model set forth in the 73 Equations
of the Manuscript.
The now highly cited 1963 paper of Pople, Santry, and Segal (Ref.~35)
comes immediately to my mind -- it included only the mathematical equations
and \textit{no numerical tests at all}.
I do include the Calculations section to show that the new model is indeed
\textit{parametrizable} and that the new terms added to the energy expression
of the model lead to great improvements in the computed molecular properties.

I have revised the Manuscript to improve its quality.
The Reviewers' comments follow below (in typewriter font) with my answers
and an account of the changes made.

\begin{verbatim}
Reviewer #1 Evaluations: 
RECOMMENDATION: Publish in JCP with mandatory revision (minor)

See attached file.
[the text extracted from file "1_reviewer_attachment_1_1311800956.pdf" follows]

Referee report on A11.06.0280

Title: A new parametrizable model of molecular electronic structure

Author: Dimitri Laikov

In this paper, a heavily parametrized semi-empirical model
is presented with significantly better accuracy than classic semi-
empirical approaches.
I find this topic should be interesting to the JCP readers and when
properly presented would deserve a publication in JCP.
However, the paper should be significantly improved before acceptance.

1. My major issue is the quality of the Section III. The section lacks
depth. This section is most important for users of the model, yet
a lot of details are not explained.
\end{verbatim}
As I noted before, the Calculations (Section III) are of less importance
than the Theory (Sections II) in this work,
but I have made some changes and additions to address this issue (see also below).

\begin{verbatim}
The only major data of the paper in Table III are poorly presented.
It took me some time to understand that these are weighted errors
multiplied by "w" factor but only after digging into the details
of the Supplementary Info knowing some of the results I guessed
the energy units are hartree. The table should be described in depth,
the units explained and all the symbols either explained in the caption
or related to the relevant section in the text.
\end{verbatim}
Footnotes are now added to Table III to explain the data,
the weighted errors
$\mathcal{E}_E$,
$\mathcal{E}_F$,
$\mathcal{E}_D$,
$\mathcal{E}_g$,
$\mathcal{E}_H$,
$\mathcal{E}_d$,
$\mathcal{E}_a$,
$\mathcal{E}_q$,
$\mathcal{E}_U$,
$\mathcal{E}_6$
in Table III were meant to be those defined in the text,
but to make it clearer one more column is now added to the table
with equation numbers for each of these terms,
the use of atomic units is stated.

\begin{verbatim}
2. Nevertheless the raw errors for the training set are still not very
useful for
a general reader. The data should be related to some other methods
and the simplest way to achieve it is to utilize some precomputed
databases. I recommend the GMTKN30 database [J. Chem. Theory Comput., 7,
291-309 (2011)]
with the data available on this webpage:
http://toc.uni-muenster.de/GMTKN/GMTKN30/GMTKN30main.html
The database is comprehensive, well organized and the
geometry information is machine readable, however,
other databases can also be used. The complete
database would be the most useful but a relevant subset covering
a wide range of chemical problems should also be sufficient. Mean errors,
mean absolute errors and maximal errors would show bias, average accuracy
and
worst case accuracy of the method and could be compared to some DFT
or wavefunction-based methods. Also, it would show how well the method
is transferable to systems not used in the training set. The errors
in the reaction energies from the Supporting Information are very good
indeed,
usually below 1 kcal/mol. Of course, some deterioration of the accuracy
is
inevitable, partly due to rather limited accuracy of the training set,
however it would show how useful is the method at this stage.
\end{verbatim}
Thanks to the Reviewer for drawing our attention
to the GMTKN30 database built up by the outstanding researchers
from the Organic Chemistry Institute
of the Westf\"alische Wilhelms-Universit\"at M\"unster, Germany.
I have visualized all 1218 molecular geometries from the GMTKN30 database --
\textit{ihre Bilder habe ich vor mein geistiges Auge gestellt}
-- and my trained eye tells me that there is likely less diversity in that set
than in my own training set of 5581 molecules!
Yes, there are some bigger molecules among those 1218,
but they are built up from some rather common fragments.
As I noted before, the Calculations (Section III) of my work were intended
mainly to show the \textit{parametrizability} of the new model
and the more extended tests should be done and reported \textit{separately},
preferably by a third party.
The first-generation parametrization of the model
reported in this work is based on the reference data
from the CCSD method with an entry-level basis set,
so I would expect the atomization energies
(and even the bond-breaking reaction energies) to agree rather poorly with any
complete-basis-set extrapolated CCSD(T) or experimental thermochemical data.
Other molecular properties -- first of all the geometries, but also
the concerted reaction energies, proton affinities,
and hopefully even the harmonic vibrational frequencies may be already
quite meaningful.

I decided to do much more than just the benchmarking of the model
on some well-established molecular database --
I provide a binary executable code of my program (in the Supplementary Information)
that can be used by anyone in the world to do molecular calculations
with the new model such as (but not limited to) the testing and benchmarking
of the model's accuracy, and I am looking forward to get any useful feedback,
either privately or in a published form.

\begin{verbatim}
3. The README in the Supporting Information should also be extended. Some
of the information in the files is self-explanatory
(e.g. errors) but I find others (e.g. some info from the files in the mol
directory)
very cryptic.
\end{verbatim}
The README file is now greatly extended to give detailed information
on the format of the data files.

\begin{verbatim}
4. The flow of the description in Section II is chaotic. Some
of the symbols are defined long after they are first used which
greatly impairs the reading. For instance, after Eq. (3),
there is a discussion of E_2 term before all the symbols in
Eq. (3) are explained. Moreover, section IIC uses some
symbols that are only defined in sections IID and IIE.
\end{verbatim}
I find no better way to write the equations of the Theory (Section II).
May the Readers forgive me this imperfection,
although I find it quite natural to explain some terms later,
this seems to me to be the best compromise.

\begin{verbatim}
5. How were the parameters optimized? With non-linear parameters, there
are
several choices of methods. How the optimization was guided not to get
stuck in a local minimum?
\end{verbatim}
I have added one sentence about the optimization algorithm to Section III.
Needless to say that for such a heavy optimization problem I have no means to make
global searches, I can only hope to be lucky enough to find the global minimum
by intuition and painful trial-and-error work, or at least to find
a good local minimum!

\begin{verbatim}
6. It would help the readers if the "lengthy but straightforward"
expressions on pages 7 and 8 were put in the main text or the
Supporting Information.
\end{verbatim}
I think that even an average undergraduate student in physical sciences
should be able to derive these expressions -- the derivatives
of the analytical formulas with only additions, subtractions, multiplications,
and divisions, however lengthy they may be.

\begin{verbatim}
7. Is the computer code described on page 8 going to be distributed?
If the author plans to distribute the code, the reference in the paper
would help the readers finding the code.
\end{verbatim}
I put a single-threaded binary executable code into the Supplementary Information,
along with an input description and sample input and output files.
Those who are interested in a deeper work with the code may try to contact me
personally for a collaboration.

\begin{verbatim}
Reviewer #2 Evaluations: 
RECOMMENDATION: Reconsider for JCP after mandatory revision (major) 
 
Reviewer #2 (Comments to the Author): 
 
The manuscripts reports an impressive amount of work that is aimed
at the development of a new general-purpose parametrizable model
for electronic structure calculations. Key ideas are the use
of a minimal basis set (MBS) of orthogonalized atom-centered basis
functions for the SCF treatment, the inclusion of polarization effects
in the spirit of second-order perturbation theory, parametrized expressions
for the one-electron and two-electron integrals (with the neglect
of many small terms), an elaborate parametrization procedure
that covers many different properties and employs CCSD reference data,
the extraction of an effective MBS Hamiltonian from CCSD calculations,
and the inclusion of dispersion effects into the two-electron integrals
via an MP2-related procedure. The parametrization is carried out
for a diverse and comprehensive basis set of 5581 molecules,
and overall error measures for the molecular properties are reported
for the recommended full model and 9 simplified variants.
The EPAPS material contains more than 879,000 lines of computer-generated
data.
 
In the Introduction, the author gives a nice general overview over the field.
In spite of the need to be selective, he should cite and briefly discuss
the work of K. F. Freed who has written more than 20 papers between 1972
and 1996 on the derivation of semiempirical MBS methods (in particular
of the MBS integrals) from rigorous ab initio theory, in a spirit that is
somewhat similar to the present work; leading references could be
Acc. Chem. Res. 1983, 16, 137 (an early review) and JCP 1996, 105, 1437
(one of the latest papers).
\end{verbatim}
Thanks to the Reviewer for making me aware of Prof. Karl~F.~Freed's work,
now I have read his key publications and I have added 4 citations of his work
(Ref.~31-34). I believe it is a good idea to cite that work,
although those model Hamiltonians are of a different kind --
to be used in a fully correlated theory (full CI or an approximation thereof).

\begin{verbatim}
The statement on page 4 that the orthogonalization-corrected methods
from ref 38-39 "did not lead to much improvement" over previous
semiempirical methods also needs to be corrected, for detailed evidence
see the following publications: JPCA 2007, 111, 5751; JCTC 2010, 6, 1546;
JCTC 2011, 7, online (dx.doi.org/10.1021/ct200434a). 
\end{verbatim}
I have now corrected that statement to "lead only to a limited improvement"
which I believe should be more diplomatic.

I have read the abstracts of the suggested publications:
\begin{enumerate}
\item \textit{Looking at Self-Consistent-Charge Density Functional Tight Binding
from a Semiempirical Perspective},
Nikolaj Otte, Mirjam Scholten, and Walter Thiel,
J. Phys. Chem. A, 2007, 111 (26), pp 5751--5755.
\item \textit{Benchmark of Electronically Excited States for Semiempirical Methods:
MNDO, AM1, PM3, OM1, OM2, OM3, INDO/S, and INDO/S2},
Mario R. Silva-Junior and Walter Thiel,
J. Chem. Theory Comput., 2010, 6 (5), pp 1546-–1564.
\item \textit{Benchmarking Semiempirical Methods for Thermochemistry, Kinetics,
and Noncovalent Interactions:
OMx Methods Are Almost As Accurate and Robust As DFT-GGA Methods
for Organic Molecules},
Martin Korth and Walter Thiel,
J. Chem. Theory Comput., Article ASAP.
\end{enumerate}
I find Paper 1 to be of little relevance to my work as it deals, first of all,
with a different type of models (SCC-DFTB) that I am not willing to criticize.
Paper 2 deals with electronically-excited states and is thus also
from a different field. Paper 3 may be indeed interesting,
although it deals only with molecules built up from 4 elements H, C, N, and O,
moreover, I have no full-text access to JCTC at my institution now,
asking a friend in the US to send me a copy of that paper would have taken some time
because he is in vacation.
Thus I cannot cite any of the 3 Papers.

\begin{verbatim}
The Theory section contains many interesting and novel ideas.
The proposed formalism clearly goes beyond the currently available
semiempirical schemes and accounts for a number of further interactions.
Having said this, one should also point out that this formalism still
employs many approximations and neglects many terms (which is mandatory
when aiming for a very fast electronic structure method). For example,
the method still uses a minimal basis, polarization is treated
at second order with some empirical damping, numerous approximations
are adopted for the molecular integrals (see II.B; use of additive schemes
and factorizations and setting small terms to zero and fitting
the remaining terms, etc), there is a fairly extensive parametrization,
and the reference ab initio methods are CCSD with a rather modest basis
(in general) and MP2 (for modeling the leading dispersion terms).
I find many of the methodological ideas interesting, but from the description
in the manuscript, it is very difficult for me to judge the soundness and
quality of the approximations in the proposed scheme. Generally speaking,
the reader would like to see more detailed justifications on the choices
and approximations made; however, I do not see how this can be done easily.
One possibility might be to give some numerical examples and integral plots
for prototypical interactions and small molecules, but I admit that this
could easily grow into a separate paper.
\end{verbatim}
I have though a lot in the last years on the foundation and justification
of parametrizable electronic structure models and made many numerical studies
on the behavior of the underlying molecular integrals that can be computed
in some way from first principles.
I decided not to go into any such details in this work
as it would have easily overwhelmed everything else to a point
that hardly anyone in the world would have ever been able to read the whole text!
The Reviewer admits rightly that "this could easily grow into a separate paper".

\begin{verbatim}
The Calculations section summarizes massive computational work
on the parametrization of 15 elements: H, Li-F, and Na-Cl.
The accuracy of the results will be limited by the use
of CCSD/TZVP-type reference data (on top of the other approximations,
see above).  The quality of the results is addressed only very summarily
by giving overall errors for various properties (Table III). This is done
in a way that baffles me.  Even if I assume that the data in Table III
are given in atomic units (not specified), I do not really understand
what the numbers mean. I do not see answers to the following simple
questions: What is the typical error of the best (full) model
for energies, geometries, dipole moments, polarizabilities, etc?
Are the errors fairly uniform for different elements or not?
How does the method perform for standard benchmark sets such as G2 and G3?
\end{verbatim}
I have improved Table III according to the comments of Reviewer \#1,
the use of atomic units is now noted.
The main point of Table III is to show how the accuracy of the model
changes as the new terms are added or removed.
The Reader can see the typical errors for energies
in the listings (file "energy.dat") found in the Supplemetary Information,
the weighted errors for all molecular properties studied
for each molecule are also listed (file "errors.dat"),
these data can be easily imported by the interested Reader
into a spreadsheet and analyzed in many ways!
Because this first-generation parametrization of the new model
is based on the reference data from CCSD calculation with an entry-level basis set,
the atomization energies would agree quite poorely with the experimental data,
but the more chemically-interesting reaction energies
should already be meaningful.
I find the reproduction of molecular geometries,
especially for weakly-bound intermolecular complexes, to be a greater challenge
than even the accurate total energies for typical molecules,
so my parametrization functional of Section IIC is constructed
to give these properties a greater boost, as discussed in the text.

To let anyone see how the model works for any molecular system of interest,
I have now included the binary executable code of my program
into the Supplementary Information, as noted above
in the answer to the comments of Reviewer \#1.

\begin{verbatim}
The EPAPS material is also essentially inaccessible to me. I appreciate
that the almost 880,000 lines of computer-generated data might contain
all the answers to my questions, but I am afraid that nobody other than
the author will be willing to check and digest this material. 
\end{verbatim}
I have now added a detailed description of the file formats so that these data
can be easily analyzed -- if not by hand then using some simple parser program
that even an average undergraduate student in physical sciences
should be able to write if needed.
To visualize all 5581 molecular structure of the training set
one can run, for example,
this command (under UNIX bash shell in an X-Window terminal)
in the \texttt{data/} directory:
\begin{verbatim}
for m in mol/*.in ; do bin/xm $m ; done
\end{verbatim}
The binary of \texttt{xm} viewer is provided in the Supplementary Information.

\begin{verbatim}
What to do with this paper? On the one hand, I see potentially interesting
ideas, but on the other hand, my dominant feeling is that I cannot really
judge the quality of the approximations made nor the quality of the results.
In my opinion, it would be a disservice to the author to publish
this potentially important manuscript as is.
\end{verbatim}
I find this statement too pessimistic.
Ideally, to "really judge the quality of the approximations made" and
"the quality of the results", an independent reproduction
of all the results of my work would be needed,
but this cannot be done quickly (in a few months)
even by the bravest worker in this field.
I fear a much greater "disservice to the author" would have been
if the Manuscript were rejected on the maximalist grounds.

\begin{verbatim}
I suggest major revision. Apart from the explicit points mentioned above,
the author should be required to improve the manuscript with regard
to the following:
(a) better justification of the approximations made;
\end{verbatim}
This point is already discussed (see above).

\begin{verbatim}
(b) better documentation of the quality of the results, including
a statistical evaluation at least with regard to the G2/G3 and S22
benchmark sets, and preferably also with regard to other established
benchmarks;
\end{verbatim}
My training set of 5581 molecules is so diverse that these benchmark sets
would hardly show anything new. Most molecules of the G2 set are already found
in my set and it is nearly so for many other benchmarks.
The traditional semiempirical models have rather few parameters
and used to be parametrized on a small training set and then
an extensive testing was very much needed to see their reliability.
The situation is very different in my case where I aim at the highest imaginable
diversity already in the training set.

\begin{verbatim}
(c) some brief information on practical matters, e.g., cpu times relative
to standard semiempirical and DFT methods;
\end{verbatim}
The CPU times would depend on the implementation,
but this model should be only marginally slower (less than twice)
than the traditional semiempirical models,
and this can be seen simply from the equations of the model,
where the bottleneck is nearly the same linear algebra
with the matrices of the same minimal-basis dimension.

\begin{verbatim}
(d) adding a detailed table of content and detailed explanation
to the EPAPS material so that readers can appreciate the documented results.
\end{verbatim}
This is done as noted above.

\vspace{1em}
I hope to have made the important changes
to improve the quality of the Manuscript.

Many thanks to everyone for the attention.


\begin{thebibliography}{00}
\bibitem{LW68}
% Consistent force field for calculations of conformations, vibrational spectra, and
% enthalpies of cycloalkane and n-alkane molecules.
S. Lifson, A. Warshel,
J. Chem. Phys. \textbf{49}, 5116 (1968).
\bibitem{WL76}
% Theoretical studies of enzymic reactions: Dielectric, electrostatic and steric
% stabilization of the carbonium ion in the reaction of lysozyme.
A. Warshel, M. Levitt,
J. Mol. Biol. \textbf{103}, 227 (1976).
\bibitem{WW80}
% An empirical valence bond approach for comparing
% reactions in solutions and in enzymes.
A. Warshel, R. M. Weiss,
J. Am. Chem. Soc. \textbf{102}, 6218 (1980).
\bibitem{B90}
% Empirical potential for hydrocarbons for use in simulating the chemical vapor
% deposition of diamond films.
D. W. Brenner,
Phys. Rev. B \textbf{42}, 9458 (1990).
\bibitem{DDLG01}
% ReaxFF: A reactive force field for hydrocarbons.
A. C. T. van Duin, S. Dasgupta, F. Lorant, W. A. Goddard III,
J. Phys. Chem. A \textbf{105}, 9396 (2001).
\bibitem{HK64}
% Inhomogeneous electron gas.
P. Hohenberg, W. Kohn,
Phys. Rev. \textbf{136}, B864 (1964).
\bibitem{H63}
% An Extended H\"uckel Theory. I. Hydrocarbons.
R. Hoffmann,
J. Chem. Phys. \textbf{39}, 1397 (1963).
\bibitem{LNV93}
% Density-matrix electronic-structure method with linear system-size scaling.
X.-P. Li, R. W. Nunes, D. Vanderbilt,
Phys. Rev. B \textbf{47}, 10891 (1993).
\bibitem{PM98}
% Canonical purification of the density matrix in electronic-structure theory.
A. H. R. Palser, D. E. Manolopoulos,
Phys. Rev. B \textbf{58}, 12704 (1998).
\bibitem{KS65}
% Self-Consistent Equations Including Exchange and Correlation Effects.
W. Kohn, L. J. Sham,
Phys. Rev. \textbf{140}, A1133 (1965).
\bibitem{MP34}
% Note on an approximation treatment for many-electron systems.
C. M{\o}ller, M. S. Plesset,
Phys. Rev. \textbf{46}, 618 (1934).
\bibitem{PCS72}
% Correlation problems in atomic and molecular systems.
% IV. Extended 	coupled-pair many-electron theory and
% its application to the BH3 molecule.
J. Paldus, J. \v{C}\'{\i}\v{z}ek, I. Shavitt,
Phys. Rev. A \textbf{5}, 50 (1972).
\bibitem{PB82}
% A full coupled-cluster singles and doubles model:
% The inclusion of disconnected triples.
G. D. Purvis, R. J. Bartlett,
J. Chem. Phys. \textbf{76} 1910 (1982).
\bibitem{RTPH89}
% A fifth-order perturbation comparison
% of electron correlation theories.
K. Ragavachari, G. W. Trucks, J. A. Pople, M. Head-Gordon,
Chem. Phys. Lett. \textbf{157}, 479 (1989).
\bibitem{B88}
% Density-functional exchange-energy approximation with correct asymptotic behavior.
A. D. Becke,
Phys. Rev. A \textbf{38}, 3098 (1988).
\bibitem{PBE96}
% Generalized gradient approximation made simple.
J. P. Perdew, K. Burke, M. Ernzerhof,
Phys. Rev. Lett. \textbf{77}, 3865 (1996).
\bibitem{B93}
% Density-functional thermochemistry. III. The role of exact exchange.
A. D. Becke,
J. Chem. Phys. \textbf{98}, 5648 (1993).
\bibitem{ITYH01}
% A long-range correction scheme for generalized-gradient-approximation
% exchange functionals.
H. Iikura, T. Tsuneda, T. Yanai, K. Hirao,
J. Chem. Phys. \textbf{115}, 3540 (2001).
\bibitem{LASCL95}
% Density functional theory including Van Der Waals forces.
B. I. Lundqvist, Y. Andersson, H. Shao, S. Chan, D. C. Langreth,
Int. J. Quantum Chem. \textbf{56}, 247 (1995).
\bibitem{ALL96}
% van der Waals interactions in density-functional theory.
Y. Andersson, D. C. Langreth, B. I. Lundqvist,
Phys. Rev. Lett. \textbf{76}, 102 (1996).
\bibitem{CP85}
% Unified approach for molecular dynamics and density-functional theory.
R. Car, M. Parrinello,
Phys. Rev. Lett. \textbf{55}, 2471 (1985).
\bibitem{DCS79}
% On first-row diatomic molecules and local density models.
B. I. Dunlap, J. W. D. Connolly, J. R. Sabin,
J. Chem. Phys. \textbf{71}, 4993 (1979).
\bibitem{L97}
% Fast evaluation of density functional exchange-correlation terms
% using the expansion of the electron density in auxiliary basis sets.
D. N. Laikov,
Chem. Phys. Lett. \textbf{281}, 151 (1997).
\bibitem{F88}
% An automatic grid generation scheme for pseudospectral self-consistent field
% calculations on polyatomic molecules.
R. A. Friesner,
J. Phys. Chem. \textbf{92}, 3091 (1988).
\bibitem{MCBRF00}
% Efficient pseudospectral methods for density functional calculations.
R. B. Murphy, Y. Cao, M. D. Beachy, M. N. Ringnalda, R. A. Friesner,
J. Chem. Phys. \textbf{112}, 10131 (2000).
\bibitem{RRS08}
% Hartree-Fock calculations with linearly scaling memory usage.
E. Rudberg, E. H. Rubensson, P. Sa\l{}ek,
J. Chem. Phys. \textbf{128}, 184106 (2008).
\bibitem{B97}
% Density-functional thermochemistry. V. Systematic optimization
% of exchange-correlation functionals.
A. D. Becke,
J. Chem. Phys. \textbf{107}, 8554 (1997).
\bibitem{L11}
% Intrinsic minimal atomic basis representation
% of molecular electronic wavefunctions.
D. N. Laikov,
Int. J. Quantum Chem. \textbf{111}, 2851 (2011).
\bibitem{ZMP94}
% From electron densities to Kohn-Sham kinetic energies, orbital energies,
% exchange-correlation potentials, and exchange-correlation energies.
Q. Zhao, R. C. Morrison, R. G. Parr,
Phys. Rev. A \textbf{50}, 2138 (1994).
\bibitem{A11}
Anonymous Reviewer of the J. Chem. Phys. (2011).
\bibitem{F72}
% A derivation of the exact pi-electron hamiltonian
K. F. Freed,
Chem. Phys. Lett. \textbf{13}, 181 (1972).
\bibitem{IF76}
% Nonclassical terms in the true effective valence shell
% Hamiltonian: A second quantized formalism.
S. Iwata, K. F. Freed,
J. Chem. Phys. \textbf{65}, 1071 (1976).
\bibitem{F83}
% Is there a bridge between ab initio and semiempirical theories of valence?
K. F. Freed,
Acc. Chem. Res. \textbf{16}, 137 (1983).
\bibitem{MF96}
% Ab initio computation of semiempirical $\pi$-electron methods. V. Geometry
% dependence of H$^v$ $\pi$-electron effective integrals.
C. H. Martin, K. F. Freed,
J. Chem. Phys. \textbf{105}, 1437 (1996).
\bibitem{PSS63}
% Approximate self-consistent molecular orbital theory. I. Invariant procedures.
J. A. Pople, D. P. Santry, G. A. Segal,
J. Chem. Phys. \textbf{43}, S129 (1963).
\bibitem{D75}
% Quantum organic chemistry
M. J. S. Dewar,
Science \textbf{187}, 1037 (1975).
\bibitem{DT77}
% Ground states of molecules. 38. The MNDO method. Approximations and parameters.
M. J. S. Dewar, W. Thiel,
J. Am. Chem. Soc. \textbf{99}, 4899 (1977).
\bibitem{NJ80}
% SINDO1.  A semiempirical SCF MO method for molecular binding energy and geometry.
% I.  Approximations and parametrization.
D. N. Nanda, K. Jug,
Theor. Chim. Acta, \textbf{57}, 95 (1980).
\bibitem{DZHS85}
% AM1: A new general purpose quantum mechanical molecular model.
M. J. S. Dewar, E. G. Zoebisch, E. F. Healy, J. J. P. Stewart,
J. Am. Chem. Soc. \textbf{107}, 3902 (1985).
\bibitem{S89}
% Optimization of parameters for semiempirical methods. I. Method.
J. J. P. Stewart,
J. Comp. Chem. \textbf{10}, 209 (1989).
\bibitem{FGZ87}
% CNDO-S2: a semiempirical SCF MO method for transition metal organometallics.
M. Y. Filatov, O. V. Gritsenko, G. M. Zhidomirov,
Theor. Chim. Acta \textbf{72}, 211 (1987).
\bibitem{S07}
% Optimization of parameters for semiempirical methods V:
% Modification of NDDO approximation and application to 70 elements.
J. J. P. Stewart,
J. Mol. Model. \textbf{13}, 1173 (2007).
\bibitem{KT93}
% Beyond the MNDO model: methodical considerations and numerical results.
M. Kolb, W. Thiel,
J. Comp. Chem. \textbf{14}, 775 (1993).
\bibitem{WT00}
% Orthogonalization corrections for semiempirical methods.
W. Weber, W. Thiel,
Theor. Chem. Acc. \textbf{103}, 495 (2000).
\bibitem{NH07}
% Semi-empirical molecular orbital methods including dispersion corrections
% for the accurate prediction of the full range of intermolecular interactions
% in biomolecules
J. P. McNamara, I. H. Hillier,
Phys. Chem. Chem. Phys. \textbf{9}, 2362 (2007).
\bibitem{TT08}
% OMx-D: semiempirical methods with orthogonalization and dispersion
% correction. Implementation and biochemical application.
T. Tuttle, W. Thiel,
Phys. Chem. Chem. Phys. \textbf{10}, 2159 (2008).
\bibitem{APS77}
% Intermolecular forces in simple systems (?)
R. Ahlrichs, R. Penco, G. Scoles,
Chem. Phys. \textbf{19}, 119 (1977).
\bibitem{RFSH98}
% Semiempirical quantum chemical PM6 method
% augmented by dispersion and H-bonding correction
% terms reliably describes various types of noncovalent
% complexes
J. \v{R}ez\'{a}\v{c}, J. Fanfrl\'{\i}k, D. Salahub, P. Hobza,
J. Chem. Theory Comput. \textbf{5}, 1749 (2009).
\bibitem{K10}
% Third-generation hydrogen-bonding corrections for
% semiempirical QM methods and force fields.
M. Korth,
J. Chem. Theory Comput. \textbf{6}, 3808 (2010).
\bibitem{RMW82}
% CNDO/2-FPP atom-in-molecule polarizabilities.
C. H. Rhee, R. M. Metzger, F. M. Wiygul,
J. Chem. Phys. \textbf{77}, 899 (1982).
\bibitem{GY05}
% Improvement of semiempirical response properties with charge-dependent
% response density.
T. J. Giese, D. M. York,
J. Chem. Phys. \textbf{123}, 164108 (2005).
\bibitem{CSG08}
% Self-consistent polarization neglect of diatomic differential overlap:
% Application to water clusters.
D. T. Chang, G. K. Schenter, B. C. Garrett,
J. Chem. Phys. \textbf{128}, 164111 (2008).
\bibitem{LH04}
% Approaching the Basis Set Limit in Density Functional Theory Calculations
% Using Dual Basis Sets without Diagonalization.
W. Liang, M. Head-Gordon,
J. Phys. Chem. A \textbf{108}, 3206 (2004).
\bibitem{SH07}
% Dual-basis self-consistent field methods: 6-31G*
% calculations with a minimal 6-4G primary basis.
R. P. Steele, M. Head-Gordon,
Mol. Phys. \textbf{105}, 2455 (2007).
\bibitem{DGG09}
% Approaching the Hartree-Fock limit by perturbative methods.
J. Deng, A. T. B. Gilbert, P. M. W. Gill,
J. Chem. Phys. \textbf{130}, 231101 (2009).
\bibitem{L76}
V. I. Lebedev,
Zh. Vychisl. Mat. i Mat. Fiz. \textbf{16}, 293 (1976).
\bibitem{LL99}
V. I. Lebedev, D. N. Laikov,
Russ. Acad. Sci. Dokl. Math. \textbf{59}, 477 (1999).
\bibitem{L05}
% A new class of atomic basis functions for accurate electronic structure
% calculations of molecules.
D. N. Laikov,
Chem. Phys. Lett. \textbf{416}, 116 (2005).
\bibitem{epaps}
See EPAPS Document No. [\textit{number will be inserted by publisher}]
for reference molecular data, parametrization control settings,
optimized parameters, error analysis, as well as the program binary
and examples of molecular calculations.
For more information on EPAPS, see http://www.aip.org/pubservs/epaps.html .
\bibitem{S96}
% Application of localized molecular orbitals to the solution of semiempirical
% self-consistent field equations.
J. J. P. Stewart,
Int. J. Quantum Chem. \textbf{58}, 133 (1996).
\bibitem{DMS97}
% Semiempirical methods with conjugate gradient density matrix search
% to replace diagonalization for molecular systems containing
% thousands of atoms.
A. D. Daniels, J. M. Millam, G. E. Scuseria,
J. Chem. Phys. \textbf{107}, 425 (1997).
\bibitem{AABBA04}
% LocalSCF method for semiempirical quantum-chemical calculation
% of ultralarge biomolecules.
N. A. Anikin, V. M. Anisimov, V. L. Bugaenko, V. V. Bobrikov, A. M. Andreyev,
J. Chem. Phys. \textbf{121}, 1266 (2004).
\bibitem{LYY96}
% Linear-scaling semiempirical quantum calculations for macromolecules.
T.-S. Lee, D. M. York, W. Yang,
J. Chem. Phys. \textbf{105}, 2744 (1996).
\bibitem{DM96}
% Semiempirical molecular orbital calculations with linear system size scaling.
S. L. Dixon, K. M. Merz, Jr.,
J. Chem. Phys. \textbf{104}, 6643 (1996).
\bibitem{DS99}
% What is the best alternative to diagonalization of Hamiltonian
% in large scale semiempirical calculations?
A. D. Daniels, G. E. Scuseria,
J. Chem. Phys. \textbf{110}, 1321 (1999).
\end{thebibliography}
\end{document}